\documentstyle[12pt]{article}

\newtheorem{theorem} {\em{Theorem}}
\newtheorem{corollary}[theorem]{\em{Corollary}}

\begin{document}
\newpage
\normalsize
\title{Quantum Codes from Cyclic Codes over $GF(4^m)$}
\author{Andrew Thangaraj, Steven W. McLaughlin}
\date{}
\maketitle
\begin{abstract}
We provide a construction for quantum codes (hermitian-self-orthogonal codes over $GF(4)$) starting from cyclic codes over $GF(4^m)$. We also provide examples of these codes some of which meet the known bounds for quantum codes. 
\end{abstract}
\section{Introduction}
In this paper we use the ideas of Calderbank et al\cite{rc} to construct a new class of quantum codes from cyclic codes over $GF(4^m)$. In particular, the following theorem from \cite{rc} can be used directly to obtain quantum codes from certain codes over $GF(4)$. 

\begin{theorem}:
\label{gfq}

Suppose $C$ is a $(n,k)$ linear code over $GF(4)$ self-orthogonal with respect to the hermitian inner product. Suppose also that the minimum weight of $C^{\perp}\setminus C$ is $d$. Then a $[[n,n-2k,d]]$ quantum code can be obtained from $C$.
\end{theorem}

The hermitian inner product between $u,v\in GF(4)^n$ is defined to be
\begin{equation}
\label{hip}
	  u.v=u_1\bar{v}_1+u_2\bar{v}_2+\cdots +u_n\bar{v}_n
\end{equation}
where $\bar{w}=w^2$ for $w\in GF(4)$.
From now on, orthogonality over $GF(4)$ will be with respect to the hermitian inner product defined above.

In this paper, we consider self-orthogonal codes over $GF(4)$ that are obtained as $4$-ary images of $4^m$-ary cyclic codes of length $n|(4^m-1)$. Binary images of self-orthogonal codes over $GF(2^m)$ have been used to obtain quantum codes in \cite{mw}. Using Theorem \ref{gfq} and working with $4$-ary images usually results in smaller dimension quantum codes at the same efficiency and minimum distance.

We begin with a discussion of $q$-ary images of cyclic codes over $GF(q^m)$. Then we present a method to obtain self-orthogonal codes over $GF(4)$ as images of codes over $GF(4^m)$.  

\section{$q$-ary images}
\subsection{Some Definitions and Notation}
The finite field $GF(q^m)$ can be considered to be a vector space of dimension $m$ over the field $GF(q)$. Let $B=\{ \beta_{1},\beta_{2},\cdots ,\beta_{m}\}$ be a basis for $GF(q^m)$ over $GF(q)$. The map {\mbox{$d_B:GF(q^m)^n\mapsto GF(q)^{mn}$}} is defined so that for ${\mathbf{u}}=\{u_1,u_2\cdots,u_n\}\in GF(q^m)^n$, 
$$d_B({\mathbf{u}})=\{u_{11},u_{21},\cdots,u_{n1},u_{12},\cdots,u_{n2},\cdots,u_{1m},\cdots,u_{nm}\}$$ 
where $u_i=\sum_{j=1}^m u_{ij} \beta_j,\: u_{ij}\in GF(q),\: 1\leq i\leq n$.	

Notice that $d_B$ is not the obvious expansion of ${\mathbf{u}}$ in terms of $B$ but a permutation of the expansion.

A linear code $C$ of length $n$ over $GF(q^m)$ is a vector subspace of $GF(q^m)^n$. The $q$-ary image of $C$ with respect to $B$ is defined to be $d_B(C)$ which is a linear code of length $mn$ over $GF(q)$.

\subsection{q-ary images of $q^{m}$-ary Cyclic Codes}
A discussion of $q$-ary images of $q^{m}$-ary codes can be found in \cite{ks,ks2,gs}. We use a description of $q$-ary images, similar to that of \cite{ks}, to study self-orthogonality and we also present an alternative proof for one of the theorems in \cite{ks}. Consider a $q^{m}$-ary cyclic code, $C$, of length $n|(q^{m}-1)$ and dimension $k$ with generator polynomial $g(x)$ of degree equal to $n-k$. So,
$$g(x)=\prod_{z\in Z}(x-\gamma^z)$$
where $\gamma\in GF(q^m)$ is a primitive $n^{\textstyle{th}}$ root of unity and $Z\subset \{0,1,\cdots,n-1\}$ is the zero set of $C$ ($|Z|=n-k$). Consider the cyclotomic cosets modulo $n$ under multiplication by $q$. Let $C_s$ be the coset with coset representative $s$. Let us define,
$$Z_q=\bigcup_{\{s\;|\;C_s\subset Z\}}C_s$$
In other words, $Z_q$ is the union of all the cyclotomic cosets that are subsets of $Z$. Let $g_q(x)=\prod_{z\in Z_q}(x-\gamma^z)$. Since $Z_q\subseteq Z$ is a union of cyclotomic cosets, $g_q(x)$ is a polynomial over $GF(q)$ and $g(x)=g_q(x)g_{q^{\prime}}(x)$. The polynomial $g_q(x)$ is hereafter referred to as the {\em{maximal factor}} of $g(x)$ over $GF(q)$. Let $B=\{ \beta_{1},\beta_{2},\cdots ,\beta_{m}\} $ be a basis for $GF(q^{m})$ over $GF(q)$ and ${\mathbf{c}}=\{c_1,c_2,\cdots,c_n\}\in C$. The codeword polynomial
	\[c(x)=\sum_{i=1}^{n} c_{i}x^{i-1} ,\ \ \ \  c_{i}\in GF(q^m)\]
is expanded in terms of the basis $B$ as
	\[c(x)=\sum_{i=1}^{n} (\sum_{j=1}^{m} c_{ij}\beta_{j}) x^{i-1}\ \ \ ,\ \ c_{ij}\in GF(q)\]
Rewriting,
	\[c(x)=\sum_{j=1}^{m} (\sum_{i=1}^{n} c_{ij}x^{i-1})\beta_{j}\]
\begin{equation} 
\label{eqn1}
	c(x)=\sum_{j=1}^{m} c_j(x)\beta_{j}\ \ \ ,\ \ c_j(x)=\sum_{i=1}^{n} c_{ij}x^{i-1}
\end{equation}

Since ${\mathbf{c}}\in C$, $d_B({\mathbf{c}})=\{c_{11},c_{12},\cdots,c_{nm}\}\in d_B(C)$. It is seen that any codeword of $d_B(C)$ is a concatenation of the coefficients of the polynomials $c_j(x)\ ,\ 1\leq j\leq m$. The next theorem further characterizes these polynomials.

Let the degree of $g_{q}(x)$ be $n-l$. The following theorem from \cite{ks} relates $g_q(x)$ and $c_j(x),\ 1\leq j\leq m$. We give an alternative proof for the theorem in \cite{ks}.
\begin{theorem}:
\label{prop-exp}

The polynomials $c_{j}(x),\ \ 1\leq j\leq m$ from (\ref{eqn1}) are codeword polynomials of the cyclic code of length $n$ generated by $g_{q}(x)$.
\end{theorem}
{\em{Proof:}}

Any codeword polynomial $c(x)$ from $C$ can be written as $c(x)=g(x)m(x)$ where\\ $m(x)\in GF(q^{m})[x]/(x^{k}-1)$. So, $c(x)=g_{q}(x)g_{q^{'}}(x)m(x)$. Expanding the coefficients of $g_{q^{'}}(x)m(x)$ in the basis, $B$, and rearranging as done before, we get $g_{q^{'}}(x)m(x)=\sum_{j=1}^{m} f_{j}(x)\beta_j$ where $f_j(x)\in GF(q)[x]/(x^{l}-1)$. So, $c(x)=\sum_{j=1}^{m} f_{j}(x)g_{q}(x)\beta_{j}$. Comparing with (\ref{eqn1}), $c_{j}(x)=g_{q}(x)f_{j}(x)$. Hence, the result.

\subsection{Self-orthogonal q-ary images}
A code is self-orthogonal if it is contained in its dual, defined with respect to some inner product. It can be shown that a code is self-orthogonal if and only if the inner product of any two codewords is equal to zero.

The following theorem can be used to obtain self-orthogonal images of cyclic codes.
\begin{theorem}:

\label{prop-selfc}
Let $C$ be a cyclic code over $GF(q^m)$ of length $n|(q^m-1)$ with generator polynomial $g(x)$ whose maximal factor over $GF(q)$ is $g_q(x)$. Let $T_m(C)$ be the cyclic code (over $GF(q)$) of length $n$ generated by $g_q(x)$. If $T_m(C)\subseteq T_m(C)^{\perp}$, $d_B(C)\subseteq d_B(C)^{\perp}$ for any basis $B$.
\end{theorem}
{\em{Proof:}}
Since the codewords of $d_B(C)$ are concatenations of codewords from $T_m(C)$, the result follows.

{\em{Note}}: During the review process, one reviewer pointed out that the above theorem generalizes (by the same proof) to any linear code over $GF(q^m)$ if we define $T_m(C)$ to be the trace of $C$ with respect to $GF(q)$\cite{ms}. In the case of cyclic codes in Theorem \ref{prop-selfc}, the trace of $C$ is exactly the cyclic code of length $n$ over $GF(q)$ generated by $g_q(x)$.

\section{Quantum codes}
\subsection{Quantum BCH codes}
To be in a position to use Theorem \ref{prop-selfc} for $q=4$, we need to determine the self-orthogonality of cyclic codes over $GF(4)$. The next theorem from \cite{rc} used mainly for the construction of quantum BCH codes\cite{gb} describes a necessary and sufficient condition for self-orthogonality of cyclic codes over $GF(4)$. 
\begin{theorem}:
\label{prop-cyclic}

A linear cyclic code over $GF(4)$ of length $n|(4^m-1)$ and generator polynomial $g(x)$ is self-orthogonal if and only if
	\[g(x)g^{\dagger}(x)\equiv 0\bmod (x^{n}-1)\]

where if $g(x)=\sum_{r=0}^{n-1} g_{r}x^{r}$

	\[g^{\dagger}(x)=GCD(\overline{g}_{0}+\sum_{r=1}^{n-1} \overline{g}_{n-r}x^{r},x^{n}-1)\]
and $\bar{g}_i=g_{i}^{2}$.
\end{theorem}

Generator polynomials of cyclic codes of length $n|(4^m-1)$ over $GF(4)$ are usually specified in terms of their zeros in $GF(4^m)$. So, it is useful to interpret Theorem \ref{prop-cyclic} in terms of the zeros(or nonzeros) of the cyclic code. 

Let $\gamma\in GF(4^m)$ be a primitive $n^{th}$ root of unity. Let $g(x)=g_0+g_1x+g_2x^2+\cdots+g_{n-1}x^{n-1}$. Assume $\gamma^z$ is a root of $g(x)$. Then,
  $$g(\gamma^z)=g_0+g_1\gamma^z+g_2\gamma^{2z}+\cdots+g_r\gamma^{(n-1)z}=0$$
$$\bar{g}_0+\bar{g}_1\gamma^{2z}+\bar{g}_2\gamma^{4z}+\cdots+g_{n-1}\gamma^{(n-1)2z}=0$$ 
$$\bar{g}_0+\bar{g}_1(\gamma^{-2z})^{n-1}+\bar{g}_2(\gamma^{-2z})^{n-2}+\cdots+\bar{g}_{n-1}(\gamma^{-2z})=0$$

So, the polynomial $\bar{g}_0+\bar{g}_1x^{n-1}+\bar{g}_2x^{n-2}+\cdots+\bar{g}_{n-1}x$ has $\gamma^{-2z}$ as a root. Since $\gamma^{-2z}$ is a root of $x^n-1$, it is seen that it is also a root of $g^{\dagger}(x)$. Since the $\dagger$ operation is an involution on the factors of $x^n-1$, it follows that if $Z\subset\{0,1,2,\cdots,n-1\}$ is the set of zeros of $g(x)$, $Z^{\dagger}=\{-2z\bmod n\;|\;z\in Z\}$ is the set of zeros of $g^{\dagger}(x)$. %Let $S=\{0,1,2,\cdots,n-1\}-Z$ be the set of nonzeros of $g(x)$. Theorem \ref{prop-cyclic} states that the code generated by $g(x)$ is self-orthogonal if and only if $Z\cup Z^{\dagger}=\{0,1,2,\cdots,n-1\}$ or equivalently $(-2S)\bmod n\subseteq Z$.
The following corollary recasts Theorem \ref{prop-cyclic} in terms of zero and nonzero sets.
\begin{corollary}:
\label{lemma-cyclic}

Let $C$ be a cyclic code over $GF(4)$ of length $n|(4^m-1)$ with zero set $Z\subset\{0,1,2,\cdots,n-1\}$ and nonzero set $S=\{0,1,2,\cdots,n-1\}\setminus Z$. If $-2S\bmod n\subseteq Z$, $C$ is self-orthogonal.
\end{corollary}

{\em{Example}} : Let $n=15$. The cyclotomic cosets modulo $n$ under multiplication by 4 are \{0\},\{1,4\},\{2,8\},\{3,12\},\{5\},\{6,9\},\{7,13\},\{10\},\{11,14\}.
So, the (15,6) cyclic code (over GF(4)) with zero set $Z=\{0,5,10,1,4,11,14,3,12\}$ (zeros in GF(16)) will be self-orthogonal since the nonzeros, $S=\{7,13,2,8,6,9\}$ are such that $-2S\bmod n\subseteq Z$. Since $S$ has four consecutive integers, the minimum distance of the dual of this code is $d=5$. So by Theorem \ref{gfq}, a $[[15,3,5]]$ quantum code may be obtained from this self-orthogonal cyclic code.

A detailed table of quantum BCH codes can be found in \cite{gb}.

\subsection{Quantum codes from $4$-ary images} 
%Let $C$ be a cyclic code over $GF(4^m)$ of length $n=4^m-1$ generated by $g(x)$ with maximal factor $g_4(x)$ over $GF(4)$. Let $C_4$ be the cyclic code over $GF(4)$ of length $n$ generated by $g_4(x)$. By Theorem \ref{prop-selfc}, it is seen that if $C_4$ is self-orthogonal with respect to the hermitian inner product(i.e $g_4(x)$ satisifes Theorem \ref{prop-cyclic}), then the 4-ary image, $d_B(C)$, is also self-orthogonal for any basis $B$. Using Theorem \ref{gfq} on $d_B(C)$, we can obtain quantum codes.

%We can use the following criterion to select the set of nonzeros, $S$, of $C$ so that $d_B(C)$ is self-orthogonal. Divide the integers $Z_n=\{0,...,n-1\}$ into the cyclotomic cosets modulo $n$ under multiplication by 4. Let $C_s$ be the coset that contains $s$. $S_4$, the set of nonzeros of $C_4$, is a union of cyclotomic cosets since $C_4$ is a BCH code and it can be seen that $S_4=\bigcup_{s\in S} C_s$. Let $Z_4$ be the set of zeros of $C_4$. If $S$ is such that $-2S_4\subseteq Z_4$, $d_B(C)$ is self-orthogonal.
   
%As an example, a $16$-ary (15,4) cyclic code with zeros $\{0,5,10,1,4,11,14,3,12,2,13 \}$ results in a maximal factor, $g_4(x)$ with zeros $\{0,5,10,1,4,11,14,3,12\}$.  So, $S=\{ 6,7,8,9\}$ and $S_4=\{ 6,9,7,13,2,8\}$. Since $g_4(x)$ generates a self-orthogonal code of length 15 over $GF(4)$, the 4-ary image is a (30,8) self-orthogonal code. Hence, a $[[30,14,5]]$ quantum code may be obtained from any $GF(4)$ image of this code. This code meets the lower bound specified in \cite{rc} for codeword length $n=30$ and message length $k=14$.

The following corollary (to Theorems \ref{prop-selfc} and \ref{prop-cyclic}) may be used for construction of cyclic codes (over $GF(4^m)$) whose $GF(4)$-images are hermitian-self-orthogonal. Let $Z_n=\{0,1,2,\cdots,n-1\}$. Divide $Z_n$ into the cyclotomic cosets modulo $n$ under multiplication by 4. Let $C_s$ be the coset represented by $s$. 

\begin{corollary}:

Let $S$ be the nonzero set of a cyclic code of length $n|(4^m-1)$ over $GF(4^m)$. For $S\subset Z_n$, let $\overline{S}=Z_n\setminus S$ and $S^c=\bigcup_{s\in S} C_s$. If $-2S^c\bmod n\subseteq \overline{S^c}$, all images of $C$ are hermitian-self-orthogonal. 
\end{corollary}
{\em{Proof}}: $S^c$ and $\overline{S^c}$ are exactly the nonzero and zero sets of $T_m(C)$. By Corollary \ref{lemma-cyclic} and Theorem \ref{prop-selfc}, the result follows. 

As an example, consider the $16$-ary (15,4) cyclic code with nonzero set $S=\{ 6,7,8,9\}$. Hence, $S^c=\{6,9,7,13,2,8\}$ and $\overline{S^c}=\{0,1,4,3,12,5,10,11,14\}$. Since $-2S^c\bmod n=\{3,12,1,4,11,14\}\subset \overline{S^c}$, all images of $C$ are self-orthogonal. Using Theorem \ref{gfq}, a $[[30,14,5]]$ quantum code may be obtained from any $GF(4)$-image of this code. This code meets the lower bound specified in \cite{rc} for codeword length $n=30$ and message length $k=14$.

In general, a $(n,k,d)$ code of length $n|(4^m-1)$ over $GF(4^m)$ with self-orthogonal images over $GF(4)$ results in a $[[mn,mn-2mk,d]]$ quantum code.

\begin{table}[ht]
\begin{center}
\begin{tabular}{|c|ccc|c|} \hline
m&n&k&d&S \\ \hline
2	&	30	&	22	&	3	&	\{1,2\}		\\
	&	30	&	18	&	4	&	\{1,2,3\}  	\\
	&	30	&	14	&	5	&	\{1,2,3,4\}	\\ \hline
3	&	189	&	177	&	3	&	\{1,2\}		\\
	&	189	&	171	&	4	&	\{1,2,3\}	\\
	&	189	&	165	&	5	&	\{1,2,3,4\}	\\
	&	189	&	159	&	6	&	\{1,2,3,4,5\}	\\
	&	189	&	153	&	7	&	\{1,2,3,4,5,6\}	\\ \hline
4	&	1020	&	1004	&	3	&	\{1,2\}		\\
	&	1020	&	996	&	4	&	\{1,2,3\}	\\
	&	\vdots	&	\vdots	&	\vdots	&	\vdots		\\
	&	1020	&	796	&	29	&	\{1,2,$\cdots$,28\}	\\ \hline
\end{tabular}
\caption{Parameters $[[n,k,d]]$ of Quantum Codes for $m=2,3,4$. $S$ is the nonzero set of the RS code over $GF(4^m)$. $n=m(4^m-1),\;\;\; k=n-2m|S|,\;\;\; d=|S|+1$}
\label{tab-codes}
\end{center}
\end{table}

Table \ref{tab-codes} is a partial list of the quantum codes that can be constructed starting with Reed-Solomon codes over $GF(4^m)$ for $m=2,3,4$.

\section{Conclusions}
We have provided a construction for quantum codes starting from cyclic codes over $GF(4^m)$.

\section*{Acknowledgement}
The authors wish to thank Dr. Eric M. Rains, AT\&T Research, for his help with interpretations in \cite{rc} and the two reviewers for providing helpful comments.

\end{document}